\RequirePackage{fix-cm}
\documentclass{article}  

\usepackage{mathptmx}  
\usepackage{newtxmath}

\usepackage{amsmath}

\newcommand\nc{\newcommand}
\nc\OMIT[1]{\relax}  
\nc\VOMIT[1]{#1}     
\nc\qph{\phantom}
\renewcommand\thanks[1]{\xdef\thankshere{#1}}  
\nc\url[1]{\texttt{
#1}}  
\nc\re[1]{(\ref{#1})}

\nc\m[1]{$#1$}
\nc\qqat{}
\nc\qqatt{& \\ \mathrel}
\nc\Fat[1]{\: #1 \:}  

\nc\lt{\mathopen{}\mathclose\bgroup\left}  \nc\rt{\aftergroup\egroup\right}
\nc\0[2]{\lt\1#1{{#2}\rt}}
\nc\1[2]{\ifcase #1{#2}\or(#2)\or[#2]\or\{#2\}\or\mathord
  <{#2}\mathord>\or\langle#2\rangle\or\lvert#2\rvert\or\lVert#2\rVert\fi}
\nc\2[2]{\mathinner{\ifcase #1{#2}\or\bigl(#2\bigr)\or\bigl[#2\bigr]\or
  \bigl\{#2\bigr\}\or\bigl<#2\bigr>\or\bigl\langle#2\bigr\rangle\or
  \bigl\lvert#2\bigr\rvert\or\bigl\lVert#2\bigr\rVert\fi}}
\nc\3[2]{\mathinner{\ifcase #1{#2}\or\Bigl(#2\Bigr)\or\Bigl[#2\Bigr]\or
  \Bigl\{#2\Bigr\}\or\Bigl<#2\Bigr>\or\Bigl\langle#2\Bigr\rangle\or
  \Bigl\lvert#2\Bigr\rvert\or\Bigl\lVert#2\Bigr\rVert\fi}}
\nc\9[2]{\ifcase #1{#2}\or\left(#2\right)\or\left[#2\right]\or\left
  \{#2\right\}\or\left\langle#2\right\rangle\or\left\langle#2\right
  \rangle\or\left\lvert#2\right\rvert\or\left\lVert#2\right\rVert\fi}
\nc\8{\2}

\nc\qtensor{\mathbf}  

\nc\overarrow[2]{\overset{\scriptscriptstyle #1}
  {\raisebox{0ex}[1.25ex][0ex]{\hbox{$#2$}}}}
\nc\ltd[1]{\overarrow{\leftarrow}{#1}}
\nc\rtd[1]{\overarrow{\rightarrow}{#1}}
\nc\nablaX{\nabla_{\qqX}}          \nc\nablax{\nabla_{\qqx}}
\nc\nablaXlt{\ltd{\nabla}_{\qqX}}  \nc\nablaxlt{\ltd{\nabla}_{\qqx}}
\nc\nablaXrt{\rtd{\nabla}_{\qqX}}  \nc\nablaxrt{\rtd{\nabla}_{\qqx}}
\nc\QQT{\text{T}}  \nc\Transp{^{\QQT}}  \nc\Transpp[1]{^{\QQT_{#1}}}
\nc\tr{\operatorname{tr}}
\nc\diad{\mathbin{\vcenter{\hbox{\boldmath{$\scriptstyle\otimes$}}}}}
\nc\qqzero{\qtensor{0}}

\nc\qb{b}  \nc\qqb{\qtensor{\qb}}
\nc\qe{e}  \nc\qqe{\qtensor{\qe}}
\nc\qf{f}  \nc\qqf{\qtensor{\qf}}
\nc\qg{g}  \nc\qqg{\qtensor{\qg}}
\nc\qh{h}  \nc\qqh{\qtensor{\qh}}
\nc\qk{k}  \nc\qqk{\qtensor{\qk}}
\nc\qt{t}  \nc\qtref{\qt_{\text{ref}}}
\nc\qu{u}  \nc\qqu{\qtensor{\qu}}
\nc\qv{v}  \nc\qqv{\qtensor{\qv}}
\nc\qw{w}  \nc\qqw{\qtensor{\qw}}
\nc\qx{x}  \nc\qqx{\qtensor{\qx}}

\nc\qB{B}            \nc\qqB{\qtensor{\qB}}
\nc\QE{\mathcal{E}}  \nc\QQE{\boldsymbol{\QE}}
\nc\qF{F}            \nc\qqF{\qtensor{\qF}}
\nc\qI{I}            \nc\qqI{\qtensor{\qI}}
\nc\qR{R}            \nc\qqR{\qtensor{\qR}}
\nc\qRic[1]{\text{Ric}_{\11{#1}}{}}
\nc\qqRic[1]{\text{\textbf{Ric}}_{\11{#1}}{}}
\nc\qRie[1]{\text{Rie}_{\11{#1}}{}}
\nc\qqRie[1]{\text{\textbf{Rie}}_{\11{#1}}{}}
\nc\qX{X}            \nc\qqX{\qtensor{\qX}}

\nc\qeps{\varepsilon}        \nc\qqeps{\boldsymbol{\upvarepsilon}}  
\nc\qchi{\chi}               \nc\qqchi{\boldsymbol{\upchi}}  
\nc\qGam{\Gamma}             \nc\qqGam{\boldsymbol\qGam}

\begin{document}

\title{Compatibility condition for the Eulerian left Cauchy--Green
deformation tensor field%
\thanks{%
The research reported in this paper and carried out at BME has been supported
by the NRDI Fund (TKP2020 NC, Grant No. BME-NCS) based on the charter of
bolster issued by the NRDI Office under the auspices of the Ministry for
Innovation and Technology,  
and by the National Research, Development and Innovation Office -- NKFIH
KH 130378.  
}
 }

\author{Tam\'as F\"ul\"op%
 \\~\\
Department of Energy Engineering, Faculty of Mechanical Engineering,
 \\
Budapest University of Technology and Economics,
 \\
Budapest, Hungary;
 \\
Montavid Thermodynamic Research Group,
Budapest, Hungary%
 }


\maketitle

 \begin{abstract}
An explicit compatibility condition formula is presented
for the Eulerian left Cauchy--Green deformation tensor field.
 It
is
 shown
to be
 the appropriate
finite-strain counterpart of
Saint-Venant's compatibility condition. The difference between the Eulerian
problem and the Lagrangian one is
highlighted.

\OMIT{
}
 \end{abstract}

\section{Introduction}  \label{seca}

The left Cauchy--Green deformation tensor field of continuum mechanics is a
quantity related to the motion of a body in a Euclidean space. The motion
defines a so-called deformation gradient tensor field, and the left and
right Cauchy--Green tensor fields are defined in terms of the deformation
gradient. Therefore, given the motion, the Cauchy--Green tensors are
determined \1 1 {and are used, for example, as the variable of a constitutive
relationship that tells the elastic stress tensor emerging in a given
arrangement/configuration of the body}.

The right Cauchy--Green tensor field is defined corresponding to the
Lagrangian description of the continuum, i.e., when the position of a given
material point at the chosen reference time is used as the space variable for
the tensor field. In parallel, for the left Cauchy--Green tensor field, a
natural setting is the Eulerian description, where, at a given time, the
position of the material point at that instant is used as the space variable.

 It is both a fundamental question \1 1 {as emphasized, e.g., in
 \cite{Acharya99}} and a practical one \1 1 {related to applications} 
whether a given tensor field can be regarded as a left---or
a right---Cauchy--Green tensor field corresponding to a motion. If the answer
is yes then we call the given Cauchy--Green tensor field \emph{compatible} \1
1 {since it is compatible with a possible motion of the continuum}. In the
small-strain or infinitesimal-strain approximation, where the deformation
gradient is not far from the unit tensor and the corresponding Cauchy--Green
tensors are also near the unit tensor \1 1 {more closely, up to linear order,
both differ from the unit tensor by the Cauchy strain tensor}, the condition
for compatibility reduces to Saint-Venant's compatibility condition
(see, e.g., \cite{fung2017classical}, Sect.~4.6). This
formula says that the Cauchy strain tensor field should have zero left+right
curl in order to stem from a motion. This condition is sufficient as long as
a simply connected and complete spatial domain is considered, which is going
to be assumed throughout this paper.
The fact that compatibility has already been discussed by Saint-Venant in
itself indicates how elementary the need is for deciding whether a given
tensor field can correspond to a continuum motion.

Without the small-strain or linearization approximation, the question is
considerably more involved. The condition for the \emph{right} Cauchy--Green
deformation tensor field has long been well-studied---see the thorough
investigation \cite{Fosdick}.

 \VOMIT{%
 \underline{The first aim} of this writing is to highlight that
the case of the \emph{left} Cauchy--Green tensor actually raises \emph{two}
questions: compatibility for the field in the \emph{Lagrangian} description
and compatibility in the \emph{Eulerian} one. Regarding the former, results
and discussions can be found in \cite{Blume89,DudaMartins95,Acharya99}. Here,
the latter case is investigated.

 Namely, as
\underline
 {the second and main aim} of this article,
a compatibility formula is determined and presented for the Eulerian left
Cauchy--Green tensor field.
 }
 \OMIT{%
Regarding the \emph{left} Cauchy--Green tensor, partial \1 1 {planar-motion}
results and discussions can be found in
\cite{Blume89,DudaMartins95,Acharya99}. As is highlighted here, the case of
the left Cauchy--Green tensor actually raises \emph{two} questions:
compatibility for the field in the \emph{Lagrangian} description and
compatibility in the \emph{Eulerian} one. Here, the latter case is
investigated: a compatibility formula is determined and presented for the
Eulerian left Cauchy--Green tensor field.
 }
At this point, it is important to emphasize that the principles behind the
formula are not novel at all: as pointed out in \cite{Fosdick}, the problem
of the Eulerian left Cauchy--Green tensor can be treated analogously to the
one of the Lagrangian right Cauchy--Green tensor. Some virtues of the form of
the here-devised compatibility condition are that
 \begin{itemize}
\item
it is given explicitly in terms of the left Cauchy--Green tensor \1 1 {not in
terms of the Christoffel symbols as intermediate quantities}, and
 \item
it contains derivatives of the left Cauchy--Green tensor, not derivatives of
its inverse like in the related Christoffel symbols.
 \end{itemize}
Correspondingly, the formula
is intended to be user-friendly for applications, for example, for
implementing it in numerical simulations in programming languages that are
fast but with the price of no symbolic manipulations. The formula can help in
validating and controlling the numerical calculations: e.g., when the left
Cauchy--Green tensor is determined from some evolution or other equation,
numerical errors make compatibility gradually violated, and finding where
violation is the most apparent can offer insight into which part of the
numerical scheme (or other aspect) could be improved.

As
 \underline
{the third aim}, the formula
 can be exhibited
as the explicit Eulerian finite-strain counterpart of Saint-Venant's one,
with which it agrees in the linear approximation, as is shown here \1 1 {as
an ``obligatory'' check}.

 \OMIT{%
It is also shown that, in the linear approximation, the obtained condition
agrees with Saint-Venant's one.
To the author's knowledge, the here-presented compatibility formula for the
Eulerian left Cauchy--Green field and its relationship to the Saint-Venant
formula have not appeared elsewhere.
 }


\section{Notations and conventions}  \label{secb}

 Throughout the paper, customary notations are used; exceptions
are where some essential aspect is to be emphasized---in those cases, the
notation helps highlighting that aspect.

Since the essence of the problem, a differential geometric question, can also
be formulated without explicitly using coordinate systems \1 1
{parametrizations}, a coordinate-free tensorial description \1 1 {see, e.g.,
\cite{ONe83b}} is
 possible.
Formulae with indices below can also be read this way, as expressions in the
abstract index notation introduced by Penrose.
Naturally, all indices can also be read as classic coordinate indices.

For the problem at hand, distinction between vectors and covectors \1 1
{elements of the dual vector space} plays an important role so upper and
lower indices are distinguished; when indices are read as coordinate indices
then this distinction is the customary one between contravariant and
covariant components \1 1 {see \cite{Haupt02}, for example}.

Motion is considered in a three-dimensional Euclidean point space \m { \QE },
where Euclidean vectors serve as the tangent vectors at each point of \m {
\QE }, in other words, all the tangent spaces are the same
vector space \m { \QQE }. Correspondingly, all cotangent spaces are also the
same, namely, the dual space of Euclidean vectors, \m { \QQE^* }. These
enable a convenient, unindexed, tensorial notation \1 1 {denoted in upright
boldface as usual} for tensors of various kind: \1 0 {tensors, cotensors and
mixed ones}. This notation also expresses the coordinate free content.
Formulae below are provided in both the tensorial and the indexed notation
since both have advantages.

In order to highlight the differential geometric essence of the problem to
discuss---and also for emphasizing the difference between vectors and
 covectors\mbox{---,}  
instead of the dot-product notation \m { \qqv \cdot \qqw } for
the scalar product of Euclidean vectors \m { \qqv } and \m { \qqw }, \m {
\qh_{KL} \qv^K \qw^L } is written, i.e., the cotensor behind \1 1 {denoted by
\m { \qqh }} is explicitly displayed. \m { \qqh } is a symmetric,
nonsingular, and positive definite cotensor, in other words, it is a Riemann
metric; \m { \QE } as a manifold, equipped with \m { \qqh }, forms a
Riemannian manifold \m { \1 1 {\QE, \qqh} }. This specific \m { \qqh } is
constant along \m { \QE } \1 1 {the notion of constantness is meaningful
since all tangent spaces along \m { \QE } are the same \m { \QQE }}.
Correspondingly, the associated Riemann curvature tensor field \m {
\qqRie{\qqh} } is zero \1 2 {\m { \qRie{\qqh} ^K_{\qph{K}LMN} = 0 }}, in
other words, this Riemannian manifold is flat. As a consequence, the Ricci
tensor field \m { \qqRic{\qqh} } is also zero \1 2 {\m { \qRic{\qqh}_{KL} = 0
}}; the Riemannian manifold at hand is Ricci-flat as well.

It is \m { \qqh } \1 2 {\m { \qh_{KL} }} and its inverse \m { \qqh^{-1} } \1
2 {\m { \qh^{KL} }} via which the customary `index raising and lowering'
\1 1 {in the coordinate-free abstract index language, the distinguished and
natural identification between Euclidean covectors and vectors} is
accomplished \1 2 {\m {\qv_K = \qh_{KL} \qv^L}, \m { \qk^M = \qh^{MN} \qk_N
}}. Here and hereafter, repeated indices involve summation \1 1 {tensorial
contraction in the abstract index language} \1 2 {Einstein convention}. Note
also that, in the indexed version, inverse is not indicated explicitly,
according to the custom.

At a reference time \m { \qtref }, a material point of the body is at a point
\m { \qX } of \m { \QE }, and, along its motion, at time \m { \qt }, it is at
a point \m { \qx } in \m { \QE }. Having a space origin chosen in \m { \QE },
\m { \qX } is described by position vector \m { \qqX } \1 2 {in index
notation: \m {\qX^K}} and \m { \qx } by vector \m {\qqx} \1 2 {\m { \qx^i }}.


Material points starting from different positions \m { \qqX } arrive at
different positions \m { \qqx }, and the motion-characterizing map
 \begin{align}  \label{@12001}
\qqchi : \qqX \mapsto \qqx = \qqchi \1 1 {\qqX}
 \end{align}
is assumed to be invertible and smooth \1 1 {smooth enough for all subsequent
considerations to hold}.

The definition of the deformation gradient tensor field \m { \qqF } is
 \begin{align}  \label{@11464}
\qF^i_{\qph{i}K} = \partial_K^{} \qchi^i ,
 \qquad 
\qqF = \qqchi \diad \nablaXlt ,
 \end{align}
where in the tensorial form \m { \diad } stands for tensorial/dyadic
product, and \m { \nablaXlt } differentiates in the variable \m { \qqX } and
acts to the left, to reflect the proper tensorial order.

The left Cauchy--Green tensor field is defined as
 \begin{align}  \label{@11702}
\qB^{ij} = \qF^i_{\qph{i}K} \qh^{KL}_{} \qF_{L}^{\qph{L}j} ,
 \qquad
\qqB = \qqF \qqh^{-1} \qqF\Transp ,
 \end{align}
with \m { \Transp } denoting the transpose. Note that, although customarily
\m {\qqh^{-1} } is omitted from the notation, i.e., the index-raising role of
\m {\qqh^{-1} } is usually not displayed explicitly, the distinction
between superscripts and subscripts \1 1 {vectors and covectors} requires
showing all appearances of \m { \qqh^{-1} } \1 1 {and, analogously, of \m {
\qqh } in index-lowering situations}.

Due to its definition \re{@11702}, \m { \qqB } is \m { \qqX }-variabled,
following from that \m { \qqF } is \m { \qqX }-variabled and \m { \qqh } is
constant. Via
 \begin{align}  \label{@12546}
\qqX = \qqchi^{-1} \1 1 {\qqx} ,
 \end{align}
quantities given in the \m { \qqX }-variable based---so-called
Lagrangian---description can be transformed to the \m { \qqx }-variable
based---so-called Eulerian---description. For example, rewriting \re{@11702}
with
 \begin{align}  \label{@13436}
\qqf \1 1 {\qqx} = \qqF \0 1 { \qqchi^{-1} \1 1 {\qqx} },
 \end{align}
one obtains
 \begin{align}  \label{@13150}
\qqb \1 1 {\qqx} = \qqB \0 1 { \qqchi^{-1} \1 1 {\qqx} } ,
 \qquad
\qqb = \qqf \qqh^{-1} \qqf\Transp ,
 \qquad
 \quad
\qb^{ij} = \qf^i_{\qph{i}K} \qh^{KL}_{} \qf_{L}^{\qph{L}j} .
 \end{align}
It is this tensor field \m { \qqb } on which the present paper focuses.

Before proceeding, a few comments becoming useful below are as follows. As a
consequence of \re{@12546}, the formula that is the counterpart to
\re{@11464} is
 \begin{align}  \label{@12992}
\qf^K_{\qph{K}i}
 = \partial_i
 { \1 1 {\qchi^{-1}}^K }
 ,
 \qquad
\qqf^{-1} \1 1 {\qqx} = \qqchi^{-1} \1 1 {\qqx} \diad \nablaxlt .
 \end{align}
Also, \m { \qqF } and \m { \qqf^{-1} } act as the Jacobi tensor \1 1 {or
matrix} of the variable transformations \re{@12001} and \re{@12546},
respectively:
 \begin{align}  \label{@14586}
\nablaXlt = \nablaxlt \, \qqF ,
 \qquad
\nablaxlt = \nablaXlt \, \qqf^{-1} ,
 \qquad \quad
\partial_K = \qF^i_{\qph{i}K} \partial_i^{} ,
 \qquad
\partial_i = \qf^K_{\qph{K}i} \partial_K^{} .
 \end{align}
At this point, it is worth noting that, since the same \m { \QQE } is the
tangent space at any point of \m { \QE }, and the same \m { \QQE^* } is the
same cotangent space at any point, the \m { \nabla } notation \1 1 {i.e., \m
{ \nablaX } in the Lagrangian description and \m { \nablax } in the Eulerian
one} has a tensorial, coordinate-free, meaning \1 1 {a.k.a. the operator \m {
\text{Grad}}, i.e., the Fr\'echet derivative}. This derivative coincides with
the differential geometric covariant derivative accompanied to the constant
metric \m { \qqh } of \m { \QE }---plain partial derivatives appear in the
indexed notation \1 2 {see \re{@14586}}.
In contrast, the other Riemann metric \m { \qqg } introduced below will be
different---the related nontrivial Christoffel symbols will be shown in
\re{@19195}.

\section{Explicit compatibility condition formula for the Eulerian
left Cauchy--Green field} \label{secc}

From \re{@11702}---considered in the variable \m {\qqx}---and
\re{@13436}--\re{@13150}, we find
 \begin{align}  \label{@13330}
\qqb^{-1} = \9 1 {\qqf\Transp}^{-1} \qqh \qqf^{-1} ,
 \qquad
\qb_{ij} = \qf_i^{\qph{i}K} \qh_{KL}^{} \qf^L_{\qph{L}j} ,
 \end{align}
or, involving \re{@12992},
 \begin{align}  \label{@13627}
\qqb^{-1} & =
\2 1 { \nablaxrt \diad \qqchi^{-1} } \qqh \2 1 {\qqchi^{-1} \diad \nablaxlt } ,
 &
\qb_{ij} & = \9 1 { \partial_i \1 1 {\qchi^{-1}}^K } \qh_{KL}
\9 1 { \partial_j \1 1 {\qchi^{-1}}^L } .
 \end{align}
(This tensor has its own names, e.g., the Finger deformation tensor
\cite{curnier1991generalized} and the left Piola deformation tensor
\cite{korobeynikov2018compatibility}, and it provides an alternative way to
introduce \m { \qqb } as the inverse of this tensor defined directly via
\re{@13627}).

One consequence that can be read off from this is that, similarly to \m {
\qqh }, \m { \qqb^{-1} } is also a symmetric, nonsingular, and positive
definite cotensor; in other words, it is a Riemann metric. Therefore, \m {
\QE } as a manifold, equipped with
 \begin{align}  \label{@14217}
\qqg = \qqb^{-1} ,  \qquad  \qg_{ij} = \qb_{ij} ,
 \end{align}
forms a Riemannian manifold.

The other consequence is that, in differential geometric language, the smooth
map \m { \qqchi^{-1} } gives rise to a pullback  of \m { \qqh } to \m { \qqg
} \1 1 {see, e.g., \cite{ONe83b}%
}. Putting these two
together, we find that \m { \qqchi^{-1} } is an isometry between the
Riemannian manifold \m { \1 1 {\QE, \qqg} } and the Riemannian manifold \m {
\1 1 {\QE, \qqh}}.

Now,
an isometry brings a zero Riemann tensor to a zero Riemann tensor.
Accordingly, although \m {\qqg} is not a constant in general, we know that it
is flat:
 \begin{align}  \label{@15073}
\qqRie{\qqg} = \qqzero ,  \qquad  \qRie{\qqg}^i_{\qph{i}jkl} = 0 .
 \end{align}
Then it also follows that it is Ricci-flat as well:
 \begin{align}  \label{@15421}
\qqRic{\qqg} = \qqzero ,  \qquad  \qRic{\qqg}_{ij}^{} = 0 .
 \end{align}

To summarize, we have that, for any \m { \qqb } that stems from some motion
\1 1 {i.e., for any \m { \qqb } that is compatible with some motion}, \m {
\qqg = \qqb^{-1} } is a flat \1 1 {hence, Ricci-flat} Riemannian metric. Note
that Ricci flatness is a technically more advantageous property than flatness
since it involves a second-order tensor rather than a fourth-order one.

Let us now consider the opposite direction. Given a symmetric, nonsingular,
and positive definite \m { \qqb
 }, can it stem from a motion?

We have just seen that Ricci flatness of \m { \qqg = \qqb^{-1} } is a minimal
necessary---and technically favourable---requirement.
 In fact, it
turns out to be a sufficient condition as well. Namely, in
two and three dimensions, Ricci flatness implies that the so-called sectional
curvature is zero \1 1 {as can be found, e.g., by analysing the topmost
formula in page 88 of \cite{ONe83b}}. Zero sectional curvature leads to
flatness \1 1 {see, e.g., \cite{ONe83b}, 3.41 Proposition}. Finally, the
Killing--Hopf theorem \1 1 {see, e.g., \cite{ONe83b}, 8.25 Corollary} ensures
that any two \1 1 {simply-connected, complete} flat Riemannian manifolds \1 0
{of the same dimension} are isometric, which in the present context says that
\m { \qqb } can stem from a motion.

Therefore, if we express Ricci flatness of \m { \qqg = \qqb^{-1} } in terms
of \m {\qqb} then we have a necessary and sufficient \1 1 {hence,
compatibility} condition for \m { \qqb }.


In terms of the Christoffel symbols of the second kind---actually, a tensor
in our case\footnote{In general, the Christoffel symbols of the second kind
corresponding to the metric of a Riemannian manifold do not form a tensor. It
is only the present particular case, where the same manifold \m { \QE } is
considered with two metric tensors \m { \qqh } and \m { \qqg }, \m { \qqh }
being a flat metric and making \m { \QE } an Euclidean affine space over \m {
\qqh }, where the Christoffel symbols of the second kind of \m { \qqg } form
a tensor from the point of view of the Euclidean affine space \m{ (\QE,
\qqh)}.}---
 \begin{align}  \label{@19195}
\qqGam & = \frac{1}{2} \qqg^{-1} \9 2 { \2 1 { \qqg \diad \nablaxlt
}\Transpp{2,3} + \2 1 { \qqg \diad \nablaxlt } - \2 1 { \nablaxrt \diad \qqg
} } ,
 &
\qGam^k_{\qph{k}ij} & = \frac{1}{2} \qg^{kl} \9 1 { \partial_i \qg_{lj} +
\partial_j \qg_{il} - \partial_l \qg_{ij} } ,
 \end{align}
where \m { {}\Transpp{2,3} } indicates transpose in the second and third
{indices} \1 1 {tensorial slots in the abstract-index language}, the Ricci
tensor reads
 \begin{align}  \label{@20474}
\qqRic{\qqg} & = \nablaxrt \cdot \qqGam - \1 1 { \tr_{1,3} \qqGam} \diad
\nablaxlt + \1 1 { \tr_{1,2} \qqGam} \qqGam - \tr_{1,3} \1 1 {\qqGam \qqGam} ,
 \\ \label{@21578}  
\qRic{\qqg}_{ij} & = \partial_k^{} \qGam^k_{\qph{k}ij} - \partial_j^{}
\qGam^k_{\qph{k}ik} + \qGam^k_{\qph{k}km} \qGam^m_{\qph{m}ij} -
\qGam^k_{\qph{k}im} \qGam^m_{\qph{m}kj} ,
 \end{align}
with \m { \tr_{1,3} } denoting contraction in the first and third indices \1
1 {tensorial slots}.
The compatibility condition for \m { \qqb }
 is
 \begin{align}  \label{@22166}
\qqRic{\qqb^{-1}} = \qqzero ,  \qquad  \qRic{\qqb^{-1}}_{ij}^{} = 0 ,
 \end{align}
which yields, when written explicitly in terms of \m { \qqb },
 \begin{equation}  \label{@22470}  \begin{split}
{
 \textstyle
 \frac{1}{4} } \Bigl\{
 \qqat  
  \tr_{1,2;3,5} \8 2 { \qqb^{-1} \8 1 { \qqb \diad \nablaxlt }
  \8 1 { \qqb \diad \nablaxlt } \diad \qqb^{-1} }
 \qqat  
- 2 \qqb^{-1} \tr_{2,4} \8 2 { \8 1 { \qqb \diad \nablaxlt }
\8 1 { \qqb \diad \nablaxlt } } \qqb^{-1}
 \qqatt  
  + \tr_{1,5;3,4} \8 2{ \qqb^{-1} \diad \qqb^{-1} \8 1 { \qqb \diad
  \nablaxlt } \8 1 { \qqb \diad \nablaxlt } }
 \qqat 
 + 2 \qqb^{-1} \tr_{1,2} \8 2 { \qqb
\8 1 { \nablaxrt \diad \nablaxrt \diad \qqb } } \qqb^{-1}
 \qqatt  
 + 2 \tr_{2,3} \8 2 { \qqb^{-1} \diad \8 1 {\nablaxrt \cdot \qqb}
 \8 1 { \nablaxrt \diad \qqb } } \qqb^{-1}
 \qqat  
  + 2 \tr_{2,4} \8 2 { \8 1 { \nablaxrt \diad \qqb } \qqb^{-1}
  \8 1 { \qqb \diad \nablaxlt } } \qqb^{-1}
 \qqatt  
+ 2  \tr_{1,2} \8 2 { \qqb^{-1} \8 1 {\qqb \diad \nablaxlt \diad \nablaxlt} }
 \qqat  
- 2 \qqb^{-1} \tr_{2,4} \8 2 { \8 1 { \qqb \diad \nablaxlt }
 \qqb \8 1 { \nablaxrt \diad \qqb } \qqb^{-1} } \qqb^{-1}
 \qqatt  
  - 3 \tr_{2,6;3,5} \8 2 { \8 1 { \nablaxrt \diad \qqb } \qqb^{-1}
  \8 1 { \qqb \diad \nablaxlt } \diad \qqb^{-1} }
 \qqat  
- 2 \8 2 { \nablaxrt \diad \8 1 { \nablaxrt \cdot \qqb } } \qqb^{-1}
 \qqatt  
  - 2 \qqb^{-1} \8 1 { \qqb \cdot \nablaxlt } \diad \nablaxlt
 \qqat  
  + 2 \tr_{2,4;3,5} \8 2 { \qqb^{-1} \diad \qqb^{-1} \8 1 {
  \qqb \diad \nablaxlt } \8 1 { \qqb \diad \nablaxlt } }
 \qqatt  
- \tr_{3,4;2,5} \8 2 { \qqb^{-1} \diad \qqb^{-1}
\8 1 { \qqb \diad \nablaxlt } \qqb \8 1 { \nablaxrt \diad \qqb } } \qqb^{-1}
  \Bigr\}
& = 0 ,
 \end{split}  \end{equation}
 \begin{equation}  \label{@24563}  \begin{split}
{
 \textstyle
 \frac{1}{4} } \bigl\{
\qb_{kl} \qb_{mj} \partial_n \qb^{lk} \partial_i \qb^{nm}  
\Fat- 2 \qb_{ik} \qb_{lj} \partial_m \qb^{nl} \partial_n \qb^{mk}  
\Fat+ \qb_{ik} \qb_{mn} \partial_l \qb^{nm} \partial_j \qb^{lk}  
 \qqatt
+ 2 \qb_{ik} \qb_{lj} \qb^{mn} \partial_m \partial_n \qb^{kl}  
\Fat+ 2 \qb_{ik} \qb_{lj} \partial_m \qb^{mn} \partial_n \qb^{kl}  
 \qqatt
+ 2 \qb_{kl} \qb_{mj} \partial_n \qb^{lm} \partial_i \qb^{nk}  
+ 2 \qb_{kl} \partial_i \partial_j \qb^{kl}  
 \qqatt
-2 \qb_{ik} \qb_{lj} \qb_{mn} \qb^{pq} \partial_p \qb^{nl} \partial_q
\qb^{km}  
-3 \qb_{kl} \qb_{mn} \partial_i \qb^{lm} \partial_j \qb^{nk}  
-2 \qb_{kj} \partial_i \partial_l \qb^{lk}  
 \qqatt
- 2 \qb_{ik} \partial_l \partial_j \qb^{kl}  
+ 2 \qb_{ik} \qb_{lm} \partial_n \qb^{mk} \partial_j \qb^{nl}  
- \qb_{ik} \qb_{lj} \qb_{mn} \qb^{pq} \partial_p \qb^{nm} \partial_q
\qb^{kl}  
 \bigr\}
 & = 0 ,
 \end{split}  \end{equation}
where, for example, \m { \tr_{1,5;3,4} } indicates tensorial contraction in
the first and fifth
 slots
and another contraction in the third and fourth
 slots.
\1 2 {\re{@24563} has been calculated by hand, and verified by the xAct
packages for Wolfram Mathematica.} Notably, \re{@22470}---or \re{@24563}---%
 has been devised to be in 
a form that contains \m {\qqb }, \m { \qqb^{-1} }, and derivatives
of \m { \qqb } but no derivatives of \m { \qqb^{-1} }. Accordingly, this form
is \1 1 {relatively} friendly for applications.

As a remark, formula \re{@22470} can
 also
be considered as a case study
for how far one can get with the index-free tensorial notation.

\section{The small-strain regime}  \label{sece}

Saint-Venant's compatibility condition says that a symmetric cotensor field
\m { \qqeps } can be expressed as the symmetric derivative of a covector
field if and only if its left+right curl is zero,
 \begin{align}  \label{@26531}
\nablaXrt \times \qqeps \1 1 {\qqX} \times \nablaXlt = \qqzero ,
 \qquad
\epsilon_{KMP}^{} \epsilon_{LNQ}^{} \partial_M^{} \partial_N^{} \qeps_{PQ}^{}
= 0 ,
 \end{align}
or, equivalently \cite{Sokolnikoff}---as is straightforward to check \1 1
{e.g., taking a contraction of the left+right curl of either of \re{@26531} and
\re{@26799} to derive the other}---, if and only if
 \begin{align}  \label{@26799}
\partial_M^{} \partial_M^{} \qeps_{KL}^{} -
\partial_K^{} \partial_M^{} \qeps_{ML}^{} -
\partial_M^{} \partial_L^{} \qeps_{KM}^{} +
\partial_K^{} \partial_L^{} \qeps_{MM}^{} = 0 ,
 \end{align}
where \m{\epsilon} is the Levi-Civita permutation symbol \1 1 {pseudotensor}.

In continuum mechanical applications, one wishes to use this condition for a
\textit{vector} field, namely, the displacement vector field
 \begin{align}  \label{@27202}
\qqu \1 1 {\qqX} = \qqchi \1 1 {\qqX} - \qqX ,
 \qquad
\qu^K = \qchi^K - \qX^K .
 \end{align}
The Euclidean structure \m { \qqh } provides an identification between
vectors and covectors \1 1 {index lowering} so \m { \qqeps } satisfying the
compatibility condition can be written as the symmetric derivative of the
covector field \m {\qu_K^{} = \qh_{KL}^{} \qu^L }, that is,
 \begin{align}  \label{@27839}
\qeps_{KL}^{} & = \frac{1}{2} \9 2 { \partial_K^{} \0 1 { \qh_{LM}^{}
\qu^{M}_{} } + \partial_L^{} \0 1 { \qh_{KM}^{} \qu^{M}_{} } } ,
&
\qqeps & = \frac{1}{2} \9 2 { \nablaXrt \diad \1 1 { \qqh \qqu } +
\1 1 { \qqh \qqu } \diad \nablaXlt } .
 \end{align}

When a version in terms of the Eulerian differentiation \m { \nablax } is
needed then vector--covector identification can be done via \m { \qqg }:
 \begin{align}  \label{@28378}
\qeps_{ij}^{} & = \frac{1}{2} \9 2 { \partial_i^{} \0 1 { \qg_{jk}^{}
\qu^{k}_{} } + \partial_j^{} \0 1 { \qg_{ik}^{} \qu^{k}_{} } } ,
&
\qqeps & = \frac{1}{2} \9 2 { \nablaxrt \diad \1 1 { \qqg \qqu } +
\1 1 { \qqg \qqu } \diad \nablaxlt } ,
 \end{align}
and actually in a different way as well:
 \begin{align}  \label{@28379}
\qeps_{ij}^{} & = \frac{1}{2} \9 2 { \0 1 { \partial_i^{} \qu^{k}_{} }
\qg_{jk}^{} + \qg_{ik}^{} \0 1 { \partial_j^{} \qu^{k}_{} } } ,
&
\qqeps & = \frac{1}{2} \9 2 { \1 1 { \nablaxrt \diad \qqu } \qqg +
\qqg \1 1 { \qqu \diad \nablaxlt } } ,
 \end{align}
The two versions \re{@28378} and \re{@28379} differ because \m { \qqg
\equiv \qqg \1 1 {\qqx} } is not constant.
Fortunately, at a material point, each of the \m { \qqeps }'s provided by
\re{@27839}, \re{@28378} and \re{@28379} agree in the linearized leading
order of \m { \qqF - \qqI }, i.e., when we are in the small-strain regime
 \begin{align}  \label{@29341}
\qqF \approx \qqI , \quad \1 7 { \qqF - \qqI }
\ll 1 ,
 \quad\quad
\Rightarrow \quad \qqg \approx \qqh , \quad \qqb \approx \qqh^{-1} .
 \end{align}

 More closely,
each of these three \m { \qqeps }'s agree in the linearized leading
order with the Almansi tensor
\m { \qqe = \frac{1}{2} \1 1 { \qqh - \qqb^{-1} } },
\,
\m {  \qe_{ij} = \frac{1}{2} \1 1 { \qh_{ij} - \qb_{ij} } }.

Now, taking the linearized leading order of
\re{@24563} yields
 \begin{align}  \label{@29975}
\partial_k^{} \partial_k^{} \qe_{ij}^{} -
\partial_i^{} \partial_k^{} \qe_{kj}^{} -
\partial_k^{} \partial_j^{} \qe_{ik}^{} +
\partial_i^{} \partial_j^{} \qe_{kk}^{} +
\1 1 {\text{higher-order terms}} = 0 .
 \end{align}
This, in light of \re{@26799}, tells that compatibility condition
\re{@24563} can be considered as the
finite-strain Eulerian counterpart of the classic small-strain compatibility
condition.

\section{Discussion}  \label{secf}

In the Eulerian description, the question of compatibility of the left
Cauchy--Green deformation tensor field admits a differential geometric
interpretation and, putting various differential geometric facts together,
leads to the question of vanishing of a two-indexed cotensor field. This
resulting condition turns out to be the Eulerian finite-strain generalization
of Saint-Venant's one applicable for small strain.

In the Lagrangian variable, the left Cauchy--Green deformation tensor field
does not have such a differential geometric background. The difference is
nicely depicted by \re{@13627}: the two derivative-related tensorial
slots/indices are the two outward \1 1 {most leftward and most rightward}
ones, enabling isometric mapping of this cotensor field from one Riemannian
manifold to another one, via the derivative map. On the other side, the
Lagrangian derivative slots/indices in \re{@11702} are inward---cf.\
\re{@11464}. This ``insulation'' prevents analogous manipulations.

This observation adds to why the left Cauchy--Green deformation tensor field
is typically found suitable for Eulerian problems while, for Lagrangian ones,
the right Cauchy--Green deformation tensor field is preferred---and,
hopefully, also to how the compatibility problem of the Lagrangian left
Cauchy--Green deformation tensor field could be successfully treated.


 \section*{Acknowledgements}
Assistance from Jos\'e M. Mart\'\i n-Garc\'\i a and P\'eter V\'an regarding
the {\it xAct} packages for {\it Wolfram Mathematica\texttrademark}
(\texttt{http://www.xact.es}) is kindly appreciated.

The author thanks Roger Fosdick for valuable comments.

 \thankshere



\begin{thebibliography}{9}



 \bibitem{Acharya99}
Acharya, A.: On compatibility conditions for the left {C}auchy--{G}reen
deformation field in three dimensions.
J. Elast. 56, 95--105 (1999).
\url{https://doi.org/10.1023/A:1007653400249}

 \bibitem{fung2017classical}
Fung, Y., Tong, P., Chen, X.: Classical and Computational Solid Mechanics.
Advanced Series In Engineering Science,
World Scientific Publishing Company, Singapore, 2nd edition (2017).

 \bibitem{Fosdick}
Fosdick, R.L.: Remarks on Compatibility. In: Eskinazi, S.
(ed.) Modern Developments in the Mechanics of Continua, pp. 109--127.
Academic Press, New York London (1966).

 \bibitem{Blume89}
Blume, J.A.: Compatibility conditions for a left {C}auchy--{G}reen strain
field.
J. Elast. 21, 271--308 (1989).
\url{https://doi.org/10.1007/BF00045780}

 \bibitem{DudaMartins95}
Duda, F.P., Martins, L.C.: Compatibility conditions for the {C}auchy--{G}reen
strain fields: solutions for the plane case.
J. Elast. 39, 247--264 (1995).
\url{https://doi.org/10.1007/BF00041840}

 \bibitem{ONe83b}
O'Neill, B.: Semi-{R}iemannian geometry with applications to 
relativity. Volume 103 of {P}ure and {A}pplied {M}athematics.
Academic Press, London (1983).

 \bibitem{Haupt02}
Haupt, P.: Continuum mechanics and theory of materials.
Springer Verlag, Berlin--Heidelberg--New York, 2nd edition (2002).
\url{https://doi.org/10.1007/978-3-662-04775-0}

 \bibitem{curnier1991generalized}
Curnier, A., Rakotomanana, L. Engineering Transactions 39(3--4),461 (1991).

 \bibitem{korobeynikov2018compatibility}
Korobeynikov, S.N., Acta Mechanica 229, 1061 (2018),
DOI:10.1007/s00707-017-1972-7.

 \bibitem{Sokolnikoff}
Sokolnikoff, I.S.: Mathematical theory of elasticity. McGraw-Hill,
New York (1956).


\end{thebibliography}
\end{document}